\documentclass[fleqn,10pt]{wlscirep}
\usepackage[utf8]{inputenc}
\usepackage[T1]{fontenc}
\usepackage{soul}
\usepackage{color}

\title{Affordable Artificial Intelligence - Augmenting Farmer Knowledge with AI}

\author[1,*]{Peeyush Kumar}
\author[1,2]{Andrew Nelson}
\author[1,3]{Zerina Kapetanovic}
\author[1]{Ranveer Chandra}
\affil[1]{Microsoft, Redmond, USA}
\affil[2]{Nelson Farms, WA, USA}
\affil[3]{University of Washington, Seattle, USA}
\affil[*]{peeyush.kumar@microsoft.com}

\begin{document}

\flushbottom
\maketitle
%
%
\thispagestyle{empty}

\section{Introduction}
It is the month of April and a farm in Eastern Washington, USA is producing wheat and lentil crops. The spring is just settling in while the low temperatures are slightly above freezing. The farmer is getting ready to spray his fields as the conditions become safe from a winter runoff and frost \cite{zheng2015frost}. The plants are significantly susceptible to certain herbicides at freezing temperatures, therefore, the farmer consults the local weather station for temperature forecasts, which is located in the closest metropolitan valley about 50 miles away from the farm. The 3 day predictions show consistent temperatures above freezing point. The farmer rents equipment and orders chemicals, and starts spraying the farm. A couple of nights the temperature in certain parts of the field drop below freezing and kills around $30\%$ of the crops. Despite the availability of weather forecasts from commercial weather stations, this is a common situation which can effect up to $50\%$ of the crops \cite{papagiannaki2014agricultural,moonmicrocilmate,zheng2015frost}. This is because the climatic parameters around the plant not only vary from the nearest weather stations but also between various regions of the farm. 

Artificial intelligence (AI) technologies are key to tackle the kind of problem presented above and many more as farmers face the challenge of feeding a growing population. The AI market is projected to be USD 1.0 billion by the end of 2020 and is estimated to reach USD 4.0 billion by 2026, at a CAGR of $25.5\%$ between 2020 and 2026. AI technologies help yield healthier crops, control pests, monitor soil and growing conditions, organize data for farmers, help with workload, and improve a wide range of agriculture-related tasks in the entire food supply chain. Additionally, as climate changes, the world becomes more connected and populations increase the burden on natural resources also increase. AI technologies are helping in a major way to create more sustainable farming practices by decreasing water wastage, overuse of chemicals on farms and conserving energy usage.

Farms produce hundreds of thousands of data points on the ground daily. Farming technique which combines farming practices with the insights uncovered in these data points using AI technology is called \textit{precision farming}. Precision farming technology augments and extends farmers' deep knowledge about their land, making production more sustainable and profitable. 

As part of the larger effort at Microsoft for empowering agricultural labor force to be more productive and sustainable, this paper presents the AI technology for predicting micro-climate conditions on the farm. Micro-climate is the accumulation of climatic parameters formed around a (approximately) homogeneous and relatively smaller region \cite{jones1993plant,rosenberg1983microclimate}. Knowledge of micro-climate and micro-climate predictions are of importance in agriculture \cite{singh2018development,cai2019research}, forestry \cite{vanwalleghem2009predicting}, architecture \cite{galaso2016influence}, urban design \cite{allegrini2017coupled}, ecology conservation \cite{zellweger2019advances}, maritime \cite{8657903} and many other domains. DeepMC predicts various micro-climate parameters with $90\%+$ accuracy at IoT sensor locations deployed in various regions across the world. 

This article presents an outline and impact of a  micro-climate prediction framework - DeepMC. This framework is based on a new deep learning approach which provides a comprehensive solution to the problem of predicting micro-climates on farms. DeepMC predicts various climatic parameters such as soil moisture, humidity, wind speed,  temperature based on the requirement over a period of 12 hours - 120 hours with varying resolution of 1hour-6hours. This article presents multiple case studies and results from live deployments of DeepMC. On average $90\%+$ accuracy is reported. 

\section{Context}\label{sec:context}
Deploying AI solutions for predicting micro-climate on farms is a challenging problem. 

First, data needs to be collected from the farm before it is processed through an AI service. 

Second, this data needs to be transferred from the location it is collected to the cloud where it is processed through the AI service forecasting micro-climate. One of the biggest challenges with deploying IoT systems for data-driven agriculture is connectivity. Since most farms are located in rural areas, there is often little to no Internet connectivity and this is crucial when it comes to enabling seamless data collection. Consider the following scenario, where we want collect data on a farm that spans thousands of acres. This would require deploying sensors across the entire farm field, which all need connectivity to convey information and in turn allow us to enable applications such as micro-climate prediction or precision irrigation. Bringing this to fruition becomes even more challenging when considering the typical farming terrain. That is, signals must be able to travel through dense crop canopy at long-distance, often with no line-of-sight. 

Third, the methodology used to build AI models which forecast micro-climates needs to be accurate, reliable for daily use, replicable across farms and adaptable for various use cases. Climatic parameters are stochastic (random process) in nature and quite challenging to model for prediction tasks on farms. 
\begin{itemize}
\item \textbf{High prediction accuracy}: Generating high accuracy results is an obvious challenge for any real world deployment of a machine learning solution. In the context of micro-climate predictions, small quantity of labelled datasets, heterogeneity of features and non-stationary of input features makes the learning problem to generate highly accurate results quite challenging. 
\item \textbf{Reliability for frequent use}: Non stationarity of the climatic time series data makes it difficult to reliably characterize the input-output relationships. Each input feature effects the output variable at a different temporal scale, for example the effect of precipitation on soil moisture is instantaneous while the effect of temperature on soil moisture is accumulated over time. 
\item \textbf{Replicable for farms across the world}: Any system for micro-climate predictions is expected to perform across various terrains, geographic and climate conditions. In practice, good quality labelled data is generally not available and even if it is accessible it is not available for every terrain, geographic or climatic conditions. Therefore, smarter techniques are required to transfer model learned in one domain to another domain with little paired labelled datasets. 
\item \textbf{Adaptable for multiple use cases}: it is also a difficult space to adapt results for multiple uses cases. Various factors influence the trend of a particular climatic parameter of interest. For example, soil moisture predictions are correlated with climatic parameters such as ambient temperature, humidity, precipitation and soil temperature \cite{hummel2001soil}; While ambient humidity is correlated with parameters - ambient temperature, wind speed and precipitation \cite{zou2017verification}. This creates a challenge for a machine learning system to accept vectors of varying dimensions as input to replicate predictions for different use cases.
\end{itemize}
Lastly, the output information needs to be presented in a way which can be consumed by the end user (producer/farmer in most cases) to aid their decision making. 

\section{Methodology}
DeepMC addresses the problems outlined above. It uses FarmBeats\cite{vasisht2017farmbeats} platform and TV White Spaces (TVWS) technology\footnote{https://rdlcom.com/tv-white-space/} to address problems of data collection, data transmission and data presentation. In addition DeepMC also utilizes the nearest available weather station forecasts to learn the relationship between various climatic parameters.

\textbf{FarmBeats}: DeepMC uses FarmBeats \cite{vasisht2017farmbeats} platform to collect climatic and soil data from multiple sensors around the world. FarmBeats is an end-to-end IoT platform for data-driven agriculture, which provides consistent data collection from various sensor types with varying bandwidth constraints. We chose the FarmBeats system for this work because of high system reliability and system availability, especially during events such as power and Internet outages caused by bad weather - scenarios that are fairly common for a farm. This data collected by FarmBeats IoT sensors is persisted in the cloud and accessed there. We also use the FarmBeats platform dashboard to deliver micro-climate predictions to the end-users using their Azure marketplace offering\footnote{https://azuremarketplace.microsoft.com/en-us/marketplace/apps/\\microsoftfarmbeats.microsoft\_farmbeats}.
\\
\textbf{TVWS technology}: The challenge of transmitting data from farm locations to a compute unit is solved by utilizing a new technology called TV White Spaces (TVWS). TVWS are unused TV spectrum that can be leveraged to extend Internet connectivity to locations that can be 10s of miles away. This technology is particularly ideal for agricultural scenarios for two reasons. First, since farms are typically located in rural areas there is a lot of unused TV spectrum available that provides large amounts of bandwidth for data transmissions (a single TV channel in the US has a 6MHz bandwidth). Second, TV spectrum spans the lower megahertz UHF and VHF bands, which is ideal for long-range communication even through dense canopy. 
\\
\textbf{Weather Station Forecasts}: Weather station forecasts are collected for training and inference through commercial weather stations. Models in DeepMC are trained and tested with various weather data providers - DarkSky\footnote{https://darksky.net/dev}, NOAA\footnote{https://www.ncdc.noaa.gov/cdo-web/webservices/v2}, AgWeatherNet\footnote{https://weather.wsu.edu/}, National Weather Service\footnote{https://www.weather.gov/documentation/services-web-api} and DTN\footnote{https://cs-docs.dtn.com/apis/weather-api/}.

\subsection{DeepMC - A deep learning based framework for micro-climate predictions}
The prediction problem is solved by using a deep learning approach by combining weather station forecasts and IoT sensor data in a special way. Each of the challenges identified in the Third point of the \ref{sec:context} are addressed.
{\bf 1) Result accuracy:} Instead of predicting the climatic parameter directly, we predict the error between the nearest commercial weather station forecast and local micro-climate forecast. This is based on the hypothesis that hyper-localization of weather station forecasts is easier to learn than learning the relationships of the predicted climatic parameter with the predictor climatic parameters from ground-up. DeepMC achieves acceptable accuracy using this design model with reported $90\%+$ (MAPE - mean absolute percentage error) accuracy across various regions in the world.\\
{\bf 2) Reliability}: In order to reliably capture varying effects of climatic data a solution needs to capture multiple trends in the data in a stationary way. DeepMC utilizes a multi scale decomposition approach to capture these effects. This approach decomposes the input signals into various scales capturing trends and details in the data and allow them to be modelled in a repeatable and reliable way.\\
{\bf 3) Replicablity:} DeepMC utilizes a specialized deep learning model, called GAN \cite{goodfellow2014generative}, to transfer learnings from source farms to target farms  around the world. 
{\bf 4) Adaptability:} All of the techniques combined together in a specialized architecture enable adaptability across multiple use cases on the farm.

\section{Impact}\label{sec:results}
DeepMC is used across many different regions of the world where FarmBeats \cite{vasisht2017farmbeats} technology is deployed. In this section, we present 3 agricultural applications which are a projection of common situations effected by weather conditions. We also show some results in comparison to common models used to solve prediction tasks.

\subsection{Scenario 1 - Spraying Herbicide: Micro-temperature predictions}
\begin{figure}
  \centering
  \includegraphics[width=0.7\linewidth]{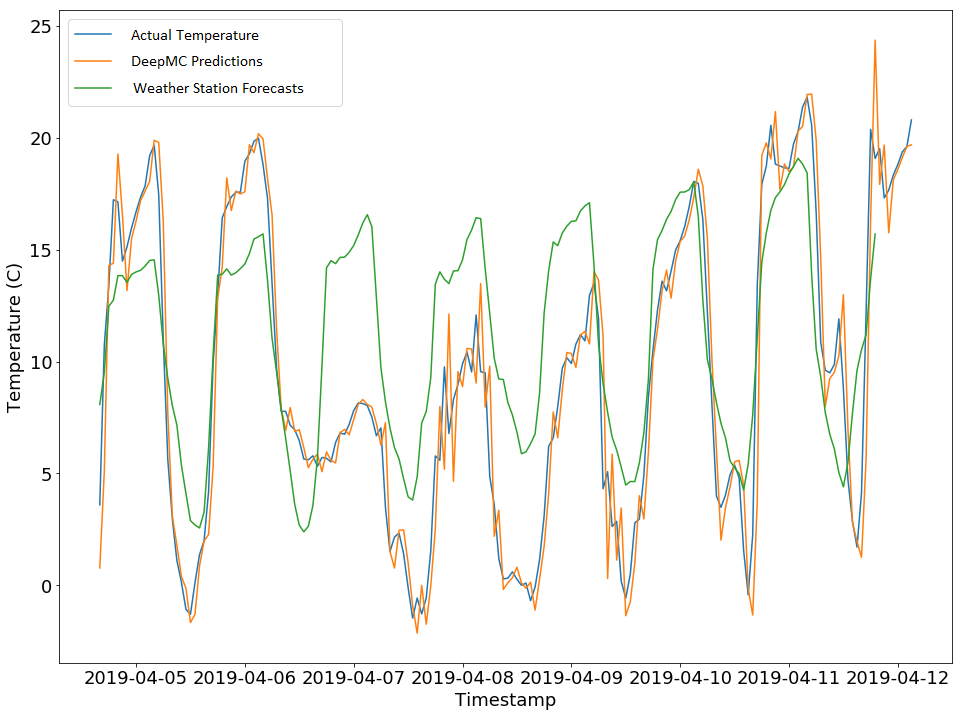}
  \caption{DeepMC Micro-Climate temperature 6 day sequential predictions with a resolution of 6 hour}
  \label{fig:temp_pred}
\end{figure}
\begin{figure}
  \centering
  \includegraphics[width=0.7\linewidth]{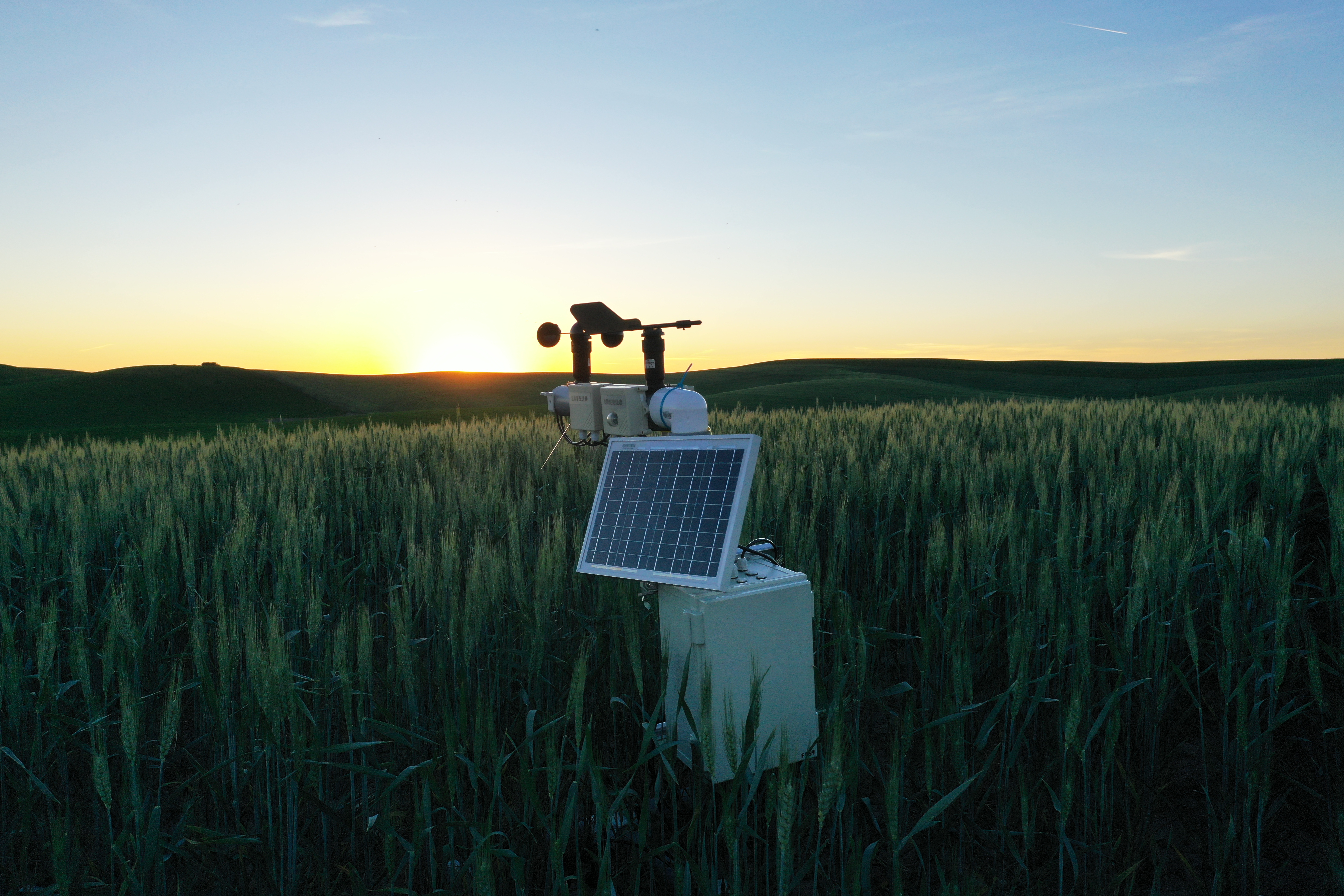}
  \caption{A FarmBeats sensor deployed on Nelson farm}
  \label{img:sensor}
\end{figure}

This scenario is the one presented in the Introduction. The farm, called Nelson Farm, is located in the eastern portion of Washington State in a region called the "Palouse".  It is an area known for it's rolling hills and crops such as wheat, lentils, peas, garbanzo beans, and canola. The area is at the foothills of mountain ranges and the combination of the rolling hills and mountains make the weather forecasts less accurate.  Many farmers own and rent land. Some of those rents are a crop share so the landlord and the farmer of the land are both affected by the decisions the farmer. There are also many research test plots scattered throughout the area that benefit from the farmer's advice on when to do certain field operations. The farmer operates on approximately 9000 acres of land across a region which is quite hilly. There are many distinct micro-climate regions in this farm. Climatic parameters vary significantly among various regions of the farm and also between the nearest commercial weather forecast provider and the readings on the ground. The farmer uses DeepMC predictions for advisory on temperature forecasts at specific locations on his farm. 
\begin{figure}
  \centering
  \includegraphics[width=0.7\linewidth]{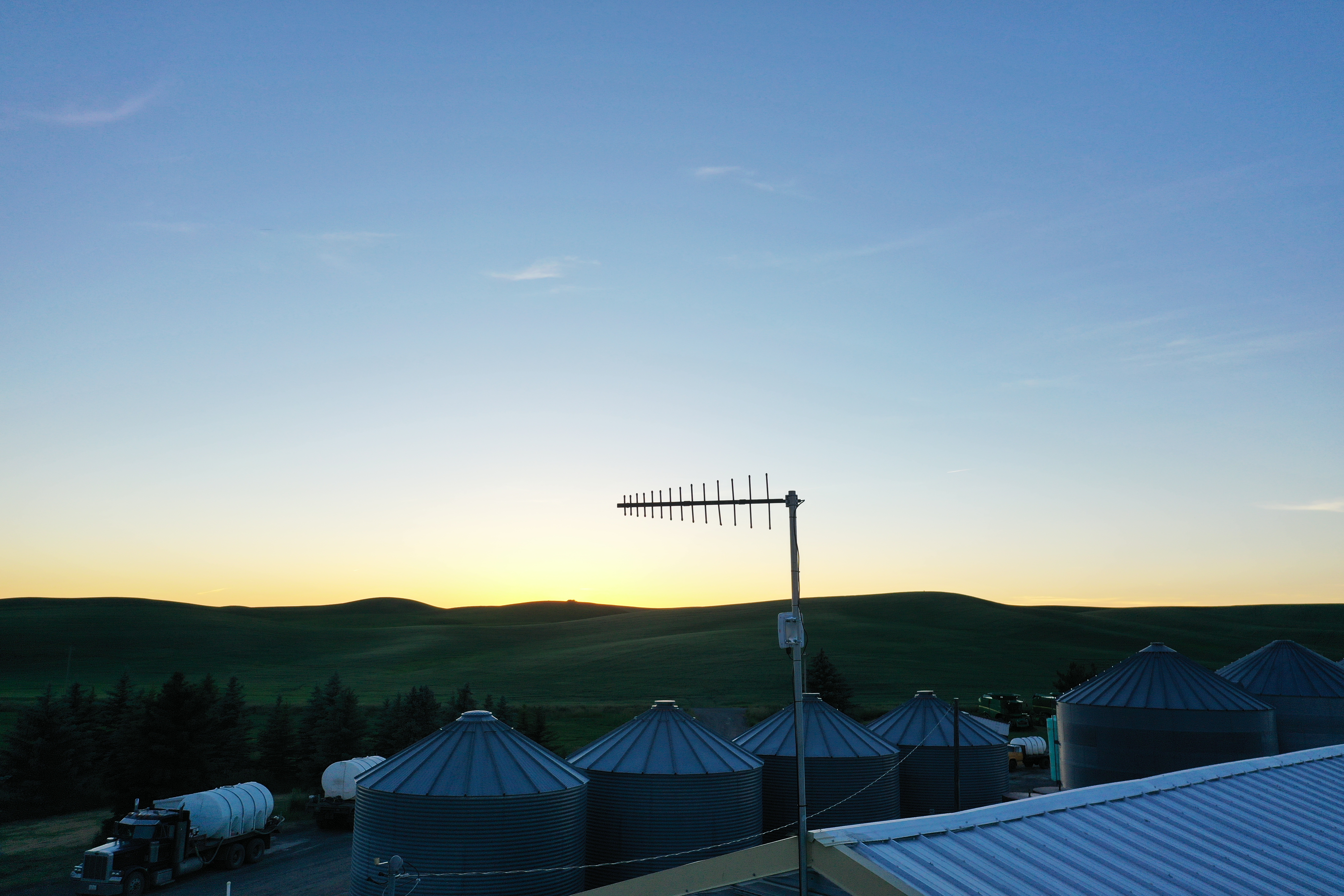}
  \caption{A TVWS deployment on Nelson farm}
  \label{img:tvws_nelson}
\end{figure}
We deployed TVWS and FarmBeats sensors at Nelson Farm. The farmer has Internet connectivity at his home, but it cannot cover the vast size of his farm that spans approximately 9000 acres across 45 miles. To remedy this, we deploy a TVWS base station that connects to the Internet at the farmers home and extends the coverage to TVWS clients that are deployed out in the farm field (see Figure \ref{img:tvws_nelson}). The TVWS links between the base station and clients are up to 13 miles in this deployment. With the TVWS deployment we were able to deploy several FarmBeats sensor boxes on the farm to collect data (see Figure~\ref{img:sensor}). These sensor boxes include many sensors such as wind speed and direction, ambient temperature, and soil moisture and temperature. 
Thus far we have had the FarmBeats system deployed on Nelson farm for 24 months and it has provided numerous insights that helped improve overall productivity. For instance in this scenario, the farmer consults DeepMC for temperature predictions for specific locations to plan logistics and operations for spraying herbicide. These experiments were conducted in the Spring of 2019 and Spring of 2020, micro-climate predictions were used daily when spraying herbicides on the wheat, lentils, peas and garbanzo beans.  It was used to plan the days that the fields could be sprayed with certain herbicides, sometimes to try to avoid freezing weather and other times to avoid overly hot weather. Figure~\ref{fig:temp_pred} shows a 6 day forecast with a temporal resolution of 6 hours. The figure shows the comparison of  the results obtained by DeepMC with Dark Sky weather forecast (from the nearest station) and the actual temperatures recorded using IoT sensors in retrospect. Based on DeepMC's predictions the farmer postponed his spraying for the period between 07-April-2019 to 11-April-2019 as the temperature predicted by DeepMC were below freezing. Instead, had the farmer relied on weather station forecasts, which consistently showed temperatures above freezing (more that 5C), then he would have been at risk of endangering the crop losing up to $30\%$ in yield. The farmer was also able to re-arrange his labor to other practices during the days that spraying herbicide wouldn't have been beneficial.  In the fall season, it is used to allow more notice to the farmer's employees on when to arrive at the farm in since many operations can not be done in freezing conditions.  This advance notice allows employees to have more time to rest when they used to have to wait for an early morning call for them to plan. In many places, especially small holder farms, this percentage is significant enough to decide whether the farmers will be able to achieve basic sustenance of food and supplies in the coming year or not. For this particular farm and location, DeepMC predictions for ambient temperature has recorded RMSE of 1.35 and MAPE of 7.68\% (implying accuracy of 92.32\%) for the data recorded in Figure~\ref{fig:temp_pred}. The predictors used for predicting micro-temperature are: a) From the IoT sensors - Ambient Temperature, Ambient Humidity, Precipitation, Wind Speed; b) From the weather station - historical ambient Temperature forecasts. The DeepMC design choice allows DeepMC architecture to adapt itself to this input-output pairing without any changes to the architecture itself.

\subsection{Scenario 2 - Phenotyping Research: Micro-soil-moisture predictions}
\begin{figure}
  \centering 
  \includegraphics[width=0.7\linewidth]{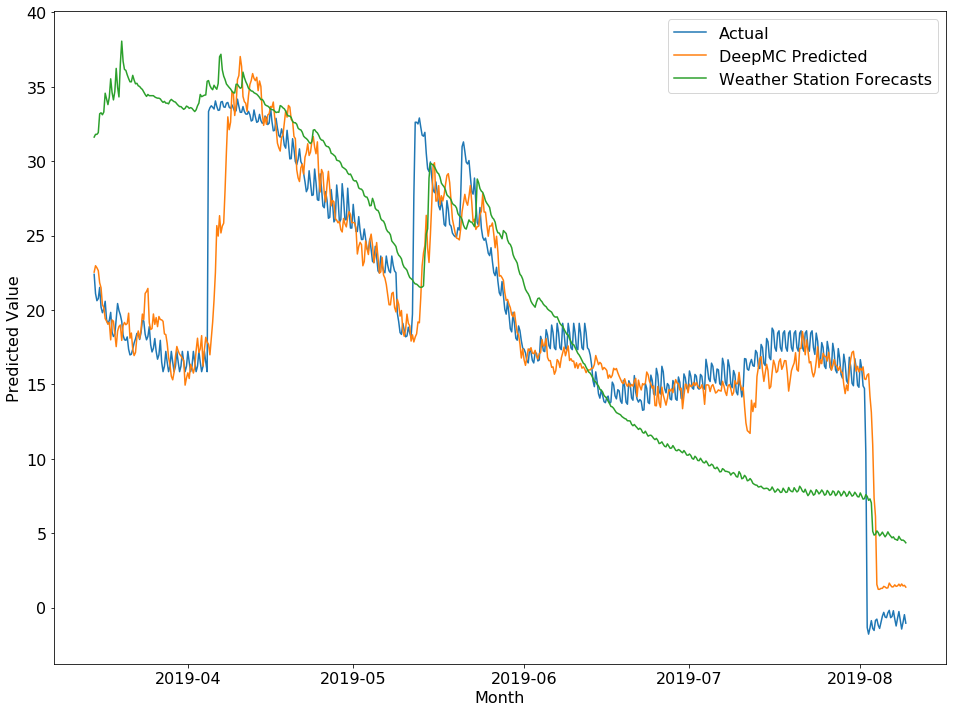}
  \caption{DeepMC Micro-climate soil moisture 4th day prediction with 6 hour resolution over a period of 10 months}
  \label{fig:sm_pred}
\end{figure}
The producer is interested in experimenting with different growing techniques for vine tomatoes. The vine tomatoes are susceptible to rot if they are too close to the soil with high moisture values. Generally, growers use trellises to lift up the vines and provide structural stability. The trellises add more challenges to manage the crops over the growing season. The producer here is interested in growing tomatoes without the trellises. This critically depends on being able to predict the local soil moisture values accurately. The producer uses DeepMC for advisory on micro-soil-moisture conditions. The results are show in Figure~\ref{fig:sm_pred} with the recorded RMSE value of 3.11 and MAPE value of 14.03\% (implying a $85.97\%$ accuracy).  The predictors used for predicting micro-soil-moisture are: a) From the IoT sensors - Ambient Temperature, Ambient Humidity, Precipitation, Wind Speed, Soil Moisture and Soil Temperature; b) From the weather station - historical Soil Moisture forecasts.
\subsection{Scenario 3 - Greenhouse control: Micro-humidity predictions}
\begin{figure}
  \centering
  \includegraphics[width=0.7\linewidth]{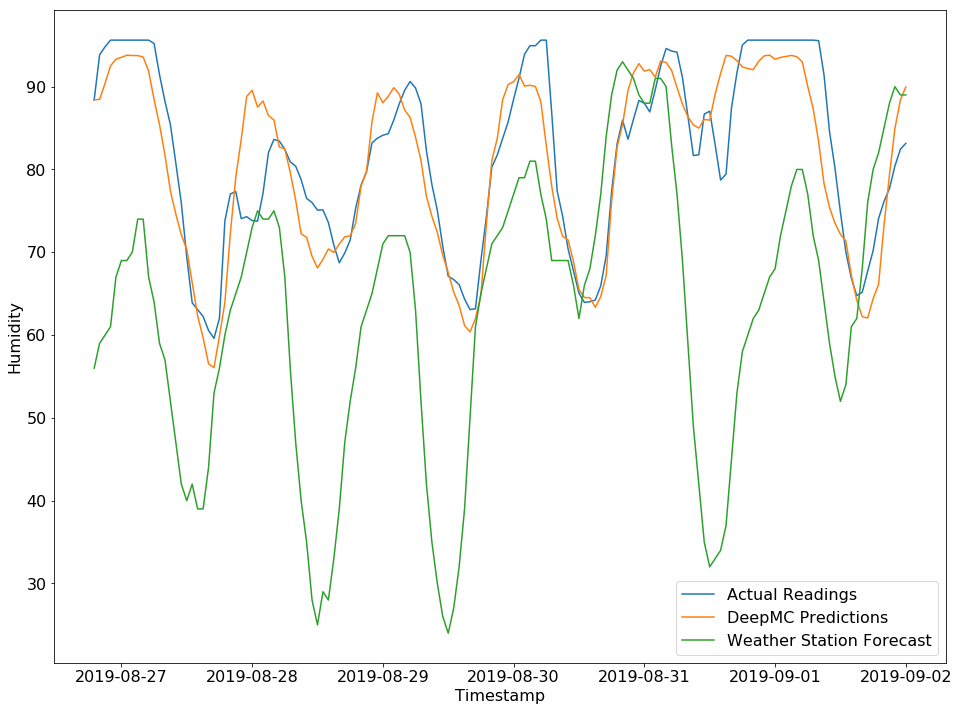}
  \caption{DeepMC Micro-Climate Humidity prediction at the 12th hour with a resolution of 1 hour over 1 week period}
  \label{fig:humidity_pred}
\end{figure}
In this scenario, the producer is storing garbanzo beans inside a grain tank. In order to control climate conditions inside the grain tank, the producer uses fans which pull the air from outside to regulate temperatures inside the greenhouse. The speed and duration of the fan control depends on the immediate humidity levels in the air outside. The producer consults DeepMC to advise in the decision making of grain tank fan control. The results are shown in Figure~\ref{fig:humidity_pred}. The predictions are plotted for the 12th hour over a period of 1 week with a resolution of 1 hour. The RMSE recorded for these predictions is 5.54 and MAPE is $5.09\%$ (therefore, the MAPE accuracy recorded is $100\%-5.09\% = 94.91\%)$. The model was trained on a stock dataset from a different farm where sufficient paired data was available and transferred at this location. The predictors used for predicting micro-humidity are: a) From the IoT sensors - Ambient Temperature, Ambient Humidity, Precipitation, Wind Speed; b) From the weather station - historical ambient Humidity forecasts.

\section{Innovation and success factors}
This work develops a comprehensive micro-climate prediction framework which can be used for multiple input-output paired climatic parameters. We highlight three real-world deployments which characterize a diverse set of conditions. We conduct a comprehensive validation of DeepMC across various regions around the world and various micro-climatic parameters. The predictions computed are being used through real-world deployments of FarmBeats system in Ireland, Africa and various states in the United States. The results presented here are computed for predictions of local temperature, local wind speed, soil moisture, soil temperature and humidity. The framework is generalizable to other input-output combinations of the  climatic parameters.

This framework is an example of how AI technologies can augment farmers' knowledge to help them make better decisions on the farm. We address some of the most difficult problems in precision agriculture viz-a-vis data collection on farms which are remote and computation to surface insights from that data. The factors which contributed to the success of this work was the innovation that went into the research along with the collaboration with the farmers who were facing the problems we set out to solve. This close link between development and application site fueled the success of this project.


\section{Sustainability}\label{sec:sus}
\textbf{Environmental Sustainability:} AI-based solutions allow for better cost control by allowing for better predictions based off of relatively affordable weather stations.  Their data allows farmers to apply chemicals with better timing allowing them to be as effective as possible.  Many weeds that are controlled by chemicals are gaining resistance, so the more effective the chemical is at the time of application, the more likely that the weed will not develop resistance to it.  This allows for less chemical application overall.  The other way that it helps with environmental issues is that farmers are able to better monitor crops that are far apart geographically and not apply the same practice at each field, instead allow the field to be managed to it's micro-climate for the current year. 

\textbf{Operational Sustainability:} We follow a partnership driven model to promote uptake of the solution described in this paper. We work closely with some of the major organizations in agriculture\footnote{For example:  https://www.businessinsider.com/microsoft-and-land-olakes-new-partnership-tackles-the-digital-divide-2020-7} (corporations, governments, cooperatives, consortiums, NGOs, etc.) who have a wider reach into the farmer ecosystem. The AI solutions are deployed in partnership with these organizations, where we share AI generated insights on market advisories, and inputs/outputs to farm operations. This creates a synergistic environment and the right incentives for organizations to deploy these solutions on the farm, where the organization benefit from the advisories (such as seed intake prediction, crop yield estimation, etc) and farmers benefit from input insights (such as micro-climate predictions, irrigation advisory, etc.). This partnership driven model has been found to scale adoptions to a wide demography of farms around the globe.

Additionally, another key requirement to make AI solutions more practical and sustainable is the cost factor. The key innovations in the technology presented here decreases the cost of deploying sensors and digital operations on the farm. The FarmBeats platform makes it cheaper to deploy and integrate sensors directly in a central datahub by its key innovations in networking technology, storage and compute on the edge framework. This central data aggregation and deployment platform also enables running AI solutions for a fraction of the cost. In addition, the PaaS model allows for scale which drives down costs per user.

A key challenge for an AI/digital framework, such as presented in this paper, to be actionable on the farm is the gap between how producers think and carry out their operations, and the complexity of using technology and digital literacy. As part of deploying the solution we also developed solutions to help educate the next generation of farmers on how technology can be utilized to make advancements in agriculture. We developed student kits, a plug-and-play platform designed to teach students about how data can be used to provide meaningful and actionable insights for farming applications. The student kit comes with several sensors (e.g. soil moisture, soil temperature, light intensity) and a Raspberry Pi running Windows IoT Core that is ready to interface with an IoT dashboard. The IoT dashboard displays all incoming sensor data that the students get to interact with, learn how to interpret data, and use the data to make intelligent decision for their farms. 

We partnered with the Future Farmers of America (FFA)\footnote{https://www.ffa.org/} to distribute FarmBeats student kits to FFA chapters across the United States. Moreover, we work with FFA to provide workshops and hackathons for teachers to learn how to educate students about ag-tech and develop lesson plans around the student kits, ultimately expanding each students view of how farming practices can become more cost-effective, productive, and sustainable by leveraging AI and IoT technology.



\section{Constraints}
There were a few challenges encountered during the development and deployment of the DeepMC framework. The development challenges are described in Section~\ref{sec:context} and how the solution presented here addresses them. Operationally, the challenges mentioned in Section~\ref{sec:sus} on Sustainability highlight the constraints on adapting this technology on farms and steps taken to overcome them.

\section{Replicability}
This framework is replicable to be able to us for other contexts. The results presented in Section~\ref{sec:results} show how DeepMC can be adapted for various farms across multiple conditions. As it stands this framework is easily scalable in a could environment.\\

DeepMC can also be used in other contexts where micro-climate predictions are needed, such as forestery, maritime environment, etc. In order to use this framework for non-farm conditions, it will be required for the model to be retrained from scratch without any change in the underlying architecture is required.

\section{Testimony}

\textbf{Andrew Nelson, Nelson Farm}: \textit{Utilizing AI allows farmers to have another tool at their disposal.  The ability to quickly apply the results that AI models produce is a great advantage.  The other large benefit is that AI can allow for other technologies to have more realized benefits. TVWS sensors that are placed throughout the farm can allow multiple predictive models for different terrain and micro-climates.  AI can has brought a new perspective on existing data, it can combine aerial imagery and ground soil moisture sensors to give insight on soil moisture that is not easy to see while walking through the farm even if the farmer were to take a soil moisture reading at multiple locations. Farmers are then able to utilize more data to make their decisions that would otherwise be difficult or too time consuming to analyze on their own.  During busy seasons, farmers are already working during all available sunlight, any time savings allows the farmer more time to tend to their crops which usually allows for higher yield potential.  The future predictions that AI provides gives farmers more insight on how to maximize their investment of time and money into the current crop.  It has allowed for larger scale testing of different farming techniques that have improved farming practices in terms of profitability, sustainability, and sometimes both.} \\
\bibliography{sample.bib}

\end{document}